\documentclass{article}

\usepackage{microtype}
\usepackage{graphicx}
\usepackage{subfigure}
\usepackage{booktabs} 

\usepackage{hyperref}

\usepackage[accepted]{icml2024}

\usepackage{amsmath}
\usepackage{amssymb}
\usepackage{mathtools}
\usepackage{amsthm}

\usepackage[capitalize,noabbrev]{cleveref}

\theoremstyle{plain}

\theoremstyle{definition}

\theoremstyle{remark}

\usepackage[textsize=tiny]{todonotes}

\icmltitlerunning{Knowledge from Large-Scale Protein Contact Prediction Models Can Be Transferred to the Data-Scarce RNA Contact Prediction Task}

\begin{document}

\twocolumn[
\icmltitle{Knowledge from Large-Scale Protein Contact Prediction Models Can Be Transferred to the Data-Scarce RNA Contact Prediction Task}



\icmlsetsymbol{corresponding}{$^{\dagger}$}

\begin{icmlauthorlist}
\icmlauthor{Yiren Jian}{corresponding,Dartmouth}
\icmlauthor{Chongyang Gao}{Northwestern}
\icmlauthor{Chen Zeng}{GWU}
\icmlauthor{Yunjie Zhao}{CCNU}
\icmlauthor{Soroush Vosoughi}{corresponding,Dartmouth}
\end{icmlauthorlist}

\icmlaffiliation{Dartmouth}{Department of Computer Science, Dartmouth College}
\icmlaffiliation{Northwestern}{Department of Computer Science, Northwestern University}
\icmlaffiliation{GWU}{Department of Physics, The Geroge Washington University}
\icmlaffiliation{CCNU}{Institute of Biophysics, Central China Normal University}

\icmlcorrespondingauthor{Yiren Jian}{yiren.jian.gr@dartmouth.edu}
\icmlcorrespondingauthor{Soroush Vosoughi}{soroush.vosoughi@dartmouth.edu}

\icmlkeywords{Machine Learning, ICML}

\vskip 0.3in
]



\printAffiliationsAndNotice{}  

\begin{abstract}

RNA, whose functionality is largely determined by its structure, plays an important role in many biological activities. The prediction of pairwise structural proximity between each nucleotide of an RNA sequence can characterize the structural information of the RNA. Historically, this problem has been tackled by machine learning models using expert-engineered features and trained on scarce labeled datasets. Here, we find that the knowledge learned by a protein-coevolution Transformer-based language model can be transferred to the RNA contact prediction task. As protein datasets are orders of magnitude larger than those for RNA contact prediction, our findings and the subsequent framework greatly reduce the data scarcity bottleneck. Experiments confirm that RNA contact prediction through transfer learning using a publicly available protein language-model is greatly improved. \emph{Our findings indicate that the learned structural patterns of proteins can be transferred to RNAs, opening up potential new avenues for research.} The code and data (for inference and training) are available for review at \url{https://anonymous.4open.science/r/ICML2024-submission-792}, which we will make public available after review.

\end{abstract}

\section{Introduction}
Proteins and RNAs are critical to many biological processes such as coding, regulation, and expression \cite{sharma2016biogenesis, goodarzi2015endogenous, esteller2011non, shi2016redefining, fire1998potent}. Understanding their structures is key to deciphering their functionalities. While experimental methods like X-ray diffraction \cite{stubbs1977structure}, nuclear magnetic resonance (NMR) \cite{cavalli2007protein}, and Cryogenic electron microscopy (Cryo-EM) \cite{glaeser2016good} can determine 3D structures, it remains challenging for structurally flexible molecules, e.g., RNAs \cite{ma2022cryo}. Consequently, the Protein Data Bank has limited RNA structures cataloged \cite{berman2000protein}.

In response, many computational tools for 3D structure prediction of biological molecules have been developed in the last decade \cite{leaver2011rosetta3, yang2015tasser, yang2020improved, kallberg2012template}. Recently, deep neural networks such as AlphaFold \cite{jumper2021highly}, ProteinMPNN \cite{dauparas2022robust}, RoseTTAFold \cite{baek2021accurate}, and Metagenomics \cite{lin2022evolutionary} have revolutionized 3D protein structure prediction, partly due to their large size and training datasets. However, this progress has not been paralleled for RNAs, mainly due to the scarcity of RNA datasets. Current RNA datasets are significantly smaller than protein datasets, with well-curated datasets containing less than 100 RNAs \cite{zerihun2021coconet} and models trained on fewer than 300 RNA structures \cite{sun2021rna, zhang2021rnacmap}. These small datasets are insufficient for training large deep neural networks, leading to RNA 3D prediction tools based on simulations (SimRNA, Rosetta FARFAR, iFoldRNA, NAST) \cite{boniecki2016simrna, das2010atomic, krokhotin2015ifoldrna, jonikas2009coarse, shi2015predicting} or fragment assembly (ModeRNA, Vfold, RNAComposer, 3dRNA) \cite{rother2011moderna, xu2014vfold, popenda2012automated, zhao2012automated, wang20193drna}.
\begin{figure*}[!t]
\centering
\includegraphics[width=0.8\textwidth]{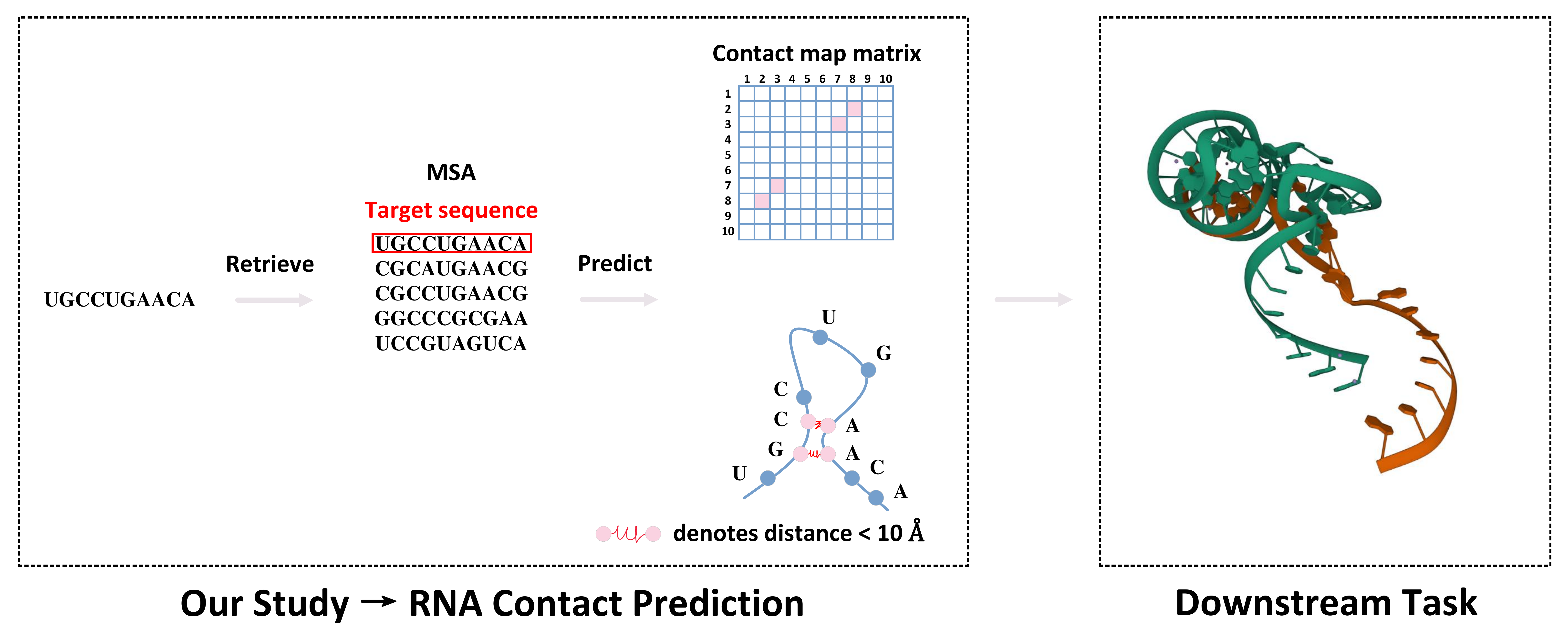}
\caption{Our study is focused on RNA contact prediction, i.e., predicting the contact map matrix for an RNA sequence. The contact map indicates the proximity between each nucleotide, with those closer than a threshold (10 \r{A}) being deemed in contact. Correct predictions of the contact map can benefit downstream tasks, e.g., by acting as constraints for filtering 3D RNA structure predictions.}
\label{fig:introduction}
\end{figure*}
In the absence of powerful 3D structure prediction models, certain structural properties of RNAs can be determined through RNA Contact Prediction \cite{jian2019direct}. The contact predictions can be used as an intermediary step to facilitate the prediction of 3D structures or directly for downstream tasks that rely on RNA structural information. For an RNA sequence of length $L$, this task aims at predicting a $L \times L$ symmetric binary matrix (called contact map) where a value of $1$ at position $(i,j)$ indicates that the $i^\text{th}$ and $j^\text{th}$ nucleotides are in contact\footnote{Contact is defined as distances smaller than a specific threshold. Following prior works, this is set by a hard distance threshold of 10 \r{A}.} with each other to each other in 3D space. The predicted contact maps capture structural constraints, which can be used for downstream tasks, such as refining RNA 3D prediction tools \cite{wang20193drna} (see Figure~\ref{fig:introduction} for an overview of the task). Note that for a target RNA sequence, the input for an RNA contact prediction model is an RNA multiple sequence alignment (MSA), which corresponds to the target RNA sequence stacked with known homologous sequences.

The first RNA Contact Prediction attempt was Direct Coupling Analysis (DCA), with variants like mfDCA \cite{morcos2011direct}, mpDCA \cite{weigt2009identification}, plmDCA \cite{ekeberg2013improved}, bmDCA \cite{muntoni2021adabmdca}, and PSICOV \cite{jones2012psicov}. These self-supervised methods infer contact maps using maximum likelihood estimation without labeled datasets. Recently, supervised methods have been explored to improve RNA contact prediction. Due to the small number of available RNA-contact map pairs, these methods rely on feature engineering, such as in RNAcontact \cite{sun2021rna}, which uses covariance matrices from Infernal \cite{nawrocki2013infernal}, PETfold-predicted secondary structures \cite{seemann2011petfold}, and RNAsol solvent accessible surface areas \cite{sun2019enhanced} to train a deep ResNet model \cite{he2016deep}. \citet{zerihun2021coconet} found that simple DCA outputs re-weighted by a convolutional layer (CoCoNet) achieve comparable RNA contact prediction precision \cite{zerihun2021coconet}.

Supervised RNA contact prediction methods leverage additional knowledge from RNA analytic tools for more informative features. These small models are necessitated by limited training examples. In contrast, abundant protein data allowed for training a large Transformer-based deep neural network, Co-evolution Transformer (CoT), for protein contact prediction~\cite{zhang2021co}. CoT was trained on 90K curated protein structures, compared to $\le 100$ RNA structures.

Since we lack the data to train such a model for RNA contact prediction from scratch, we investigate the possibility of re-using and tuning the learned parameters of a pre-trained protein language-model (such as CoT) to create an RNA contact prediction model, a process referred to as transfer learning. Inspired by recent breakthroughs in unified vision-language models \cite{bao2022vlmo, wang2022git, li2021align} and transfer learning across text and visual domains \cite{lu2021pretrained, jian-etal-2022-non}, which have demonstrated the effectiveness of transferring knowledge between related modalities, such as leveraging the structural abilities learned from code and music to enhance language models \cite{papadimitriou2020learning}, we propose that bio-molecule contact patterns learned by the CoT protein Transformer network could be transferred to improve RNA contact prediction performance.

Similar to RNA contact prediction models, CoT takes protein MSAs as input. The input to CoT is represented using English characters, with each amino acid represented by a unique English character. CoT then utilizes the attention mechanism of Transformers~\cite{vaswani2017attention} to learn the contacts, analogous to how Transformer-based language models, such as GPT-3~\cite{brown2020language}, learn dependencies between words in a given text. Though at the surface level, RNA and protein sequence data are comprised of different building blocks (nucleotides for RNAs and amino acids for proteins), we speculate that they share deeper similarities concerning their contact patterns, analogous to two languages with different lexicons but a similar syntax. Hence, it may be possible to transfer knowledge about contact patterns from one to another, analogous to cross-lingual transfer in Transformer-based language models \cite{gogoulou-etal-2022-cross}.

\begin{figure}[!t]
\centering
\includegraphics[width=0.45\textwidth]{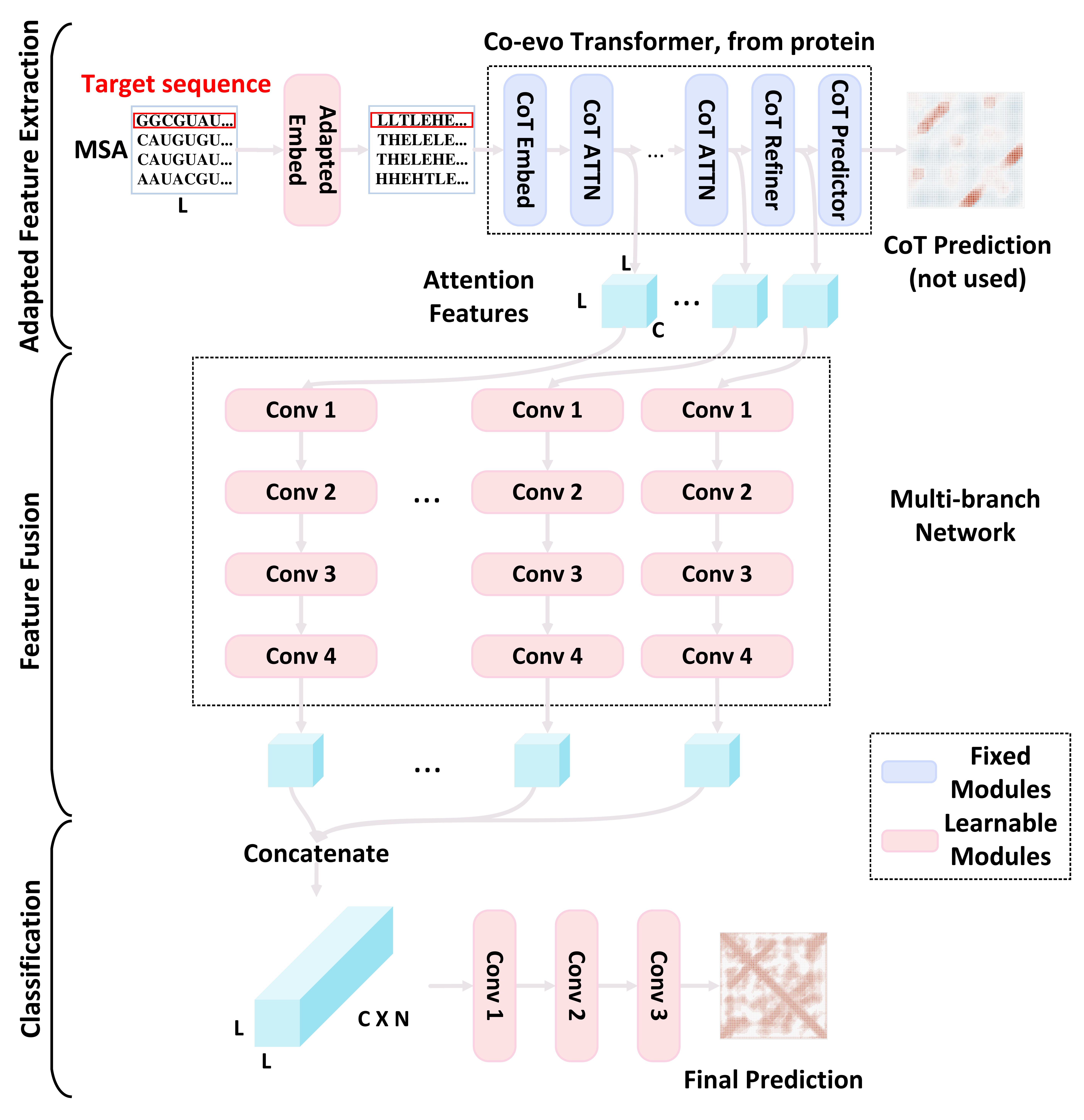}
\caption{Overview of our three-stage method (from top to bottom). \emph{Adapted Feature Extraction}: First, a projection layer is used to translate the RNA MSA sequences into protein language (e.g., from nucleotide ``AUCG'' to amino acids ``HETL''). Then, we leverage a fixed large-scale pre-trained protein contact prediction transformer model (called Co-evolution Transformer model (CoT)) to extract attentive (i.e., contribution) features at different layers. \emph{Feature Fusion}: Features from different layers are processed by separate convolution blocks before being concatenated. \emph{Classification}: The aggregated features are sent into a standard Convolutional Network (ConvNet) classifier with three layers of convolution.}
\label{fig:overview}
\end{figure}

We investigate our hypothesis by adapting the pre-trained CoT to our RNA dataset and using the adapted representations to train a convolutional network (ConvNet) for RNA contact prediction (see Figure~\ref{fig:overview} for an overview of our method). Our explorations show that this simple method, which does not rely on any additional pre-processing or feature engineering and can detect true contacts missed by prior works. In addition to improving RNA contact prediction by using knowledge from a pre-trained protein language-model, \emph{\textbf{more importantly, our study serves as a strong proof of concept for the possibility of transfer learning between the proteins and RNAs.}}

\section{Background and Related Works}
\paragraph{Unsupervised Contact Prediction Based on the Co-Evolution Hypothesis}
The \emph{co-evolution hypothesis} is the basis of many contact prediction methods (for both proteins and RNA). The hypothesis suggests that spatially proximate pairs of amino acids or nucleotides tend to co-evolve to maintain their structure and function \cite{weigt2009identification}.

In practice, this is used for RNA (and protein) contact prediction as follows: to predict the contacts of a target RNA sequence, first, a sequence database is used to find similar sequences. These sequences are likely homologs of the target, with differences due to mutations during evolution. These sequences are then aligned, creating what is called a multiple sequence alignment (a.k.a MSA). Based on the co-evolution hypothesis, the contact prediction for the target sequence can then be reformulated as detecting the co-evolution nucleotide pairs in the MSA. For example, \citet{morcos2011direct} calculate the covariance between each pair of nucleotides, thus creating a covariance matrix as an approximation of the co-evolution. The direct coupling score (DCA score) between each pair can then be computed through different approximation methods. While \citet{morcos2011direct} use mean-field approximation (mfDCA), other DCA variants (e.g., \cite{ekeberg2013improved}) use different tricks for the approximation. 

DCA methods are all purely unsupervised, based on the counting frequency of the residues in MSA, and can be applied to proteins and RNA sequences. Recently, it has been shown that transformer-based protein language-models can also be unsupervised protein contact learners \cite{rao2021transformer, rao2021msa}, though these methods are not necessarily based on the co-evolution hypothesis.

\paragraph{Supervised Contact Prediction}
Given $\sim180$K known protein structures in PDB, \citet{zhang2021co} train a 20M-parameters attention-based Transformer model for end-to-end prediction of protein contacts based on MSA. The attention mechanism of the model, called the Co-evolution Transformer (CoT), is specifically designed to model co-evolution by considering the outer product of representations of two positions. Such a model is only successful given a large labeled dataset of known MSA to contact mappings.

Training such a large model for RNA is unfortunately not practical as there are currently no such large datasets available. The largest of such data is at least 2-3 orders of magnitude smaller than what is available for proteins. To overcome this bottleneck, most resort to feature engineering to train smaller models. For instance, recent works \cite{zhang2021rnacmap, sun2021rna} combine DCA outputs (or similarly, covariance matrices) with other features (such as predicted secondary structures, solvent surface areas, etc.) extracted from different RNA analysis tools to train relatively small convolution networks, using only hundreds of labeled data point. Then, \citet{zerihun2021coconet} propose CoCoNet, showing that the output of DCA by itself is sufficient for training such models and that oftentimes expensive additional feature extraction is not needed. Finally, a recent work \cite{taubert2023rna} investigates self-supervised training with regression for RNA contact prediction.

\paragraph{Transfer Learning}
We show in this paper that the learned knowledge of a pre-trained protein contact prediction model can be effectively used for RNA contact prediction, not only removing the need for additional feature engineering and extraction but also vastly outperforming CoCoNet. Our proposed method is built upon the concept of ``Transfer Learning'' \cite{donahue2014decaf}, which assumes that knowledge learned from one task is beneficial to other related tasks. Transfer learning has enabled the adaption of large pre-trained deep neural networks to new tasks with a limited number of labeled examples. This is typically done by training newly initialized layers at the end of the pre-trained network (which tends to be task-specific) using the small dataset while keeping the other layers frozen (which preserves the learned knowledge from the previous task). Only the relatively small set of parameters in the final layers will be updated, which will adapt the network to the new task.

Transfer Learning has been shown to be effective for class-level transfer in a single domain \cite{wah2011caltech} and for different domains (e.g., an image classification model adapted for semantic segmentation \cite{long2015fully}). There is also research that shows that seemingly unrelated tasks can also help each other \cite{meyerson2021traveling}. Even models for different modalities can be transferred. For instance, \citet{mokady2021clipcap} and \citet{lu2021pretrained} show that semantic knowledge can be transferred between language and visual models.

A key challenge of RNA contact prediction is the small dataset size, which prohibits us from learning a deep model from scratch. We hypothesize (and later verify) that knowledge could be effectively transferred from a pre-trained protein contact Transformer to RNA contact prediction, enabling us to train high-performing RNA contact prediction models without the need for additional labeled or feature engineering, both of which can be prohibitively expensive. Analogies can be drawn between our approach and research done on the cross-lingual transfer of language models \cite{gogoulou-etal-2022-cross} that adapt a pre-trained model to a new language by learning its syntax while retaining the semantic knowledge in the pre-trained model; here we are adapting a biological model pre-trained on the ``language of proteins'' to the ``language of RNAs''.

\section{Methods and Setups}\label{section:experimental-setup}
\subsection{Protein-to-RNA Transferred Contact Prediction Model}
In this section, we provide details of our model's architecture, input, and output. An overview of our approach is visualized in Figure~\ref{fig:overview}. 

\subsubsection{MSA as Input}\label{Methods:input}
Our RNA contact prediction model relies on the CoT model, which takes protein MSA as the input. Thus, we need to adapt or map the RNA language, which is comprised of nucleotides, to the protein language, which is comprised of amino acids. Specifically, suppose our target RNA MSA has $M$ aligned sequences, each with the length of $L$ nucleotides. Then, the RNA MSA can be represented as a $M \times L$ matrix, with each element being  ``A'', ``U'', ``C''', ``G'', ``-'', where ``-'' denotes a gap in the alignment. As the CoT embedding layer recognizes only symbols corresponding to amino acids and not nucleotides, we assign each type of nucleotide in the RNA MSA to an amino acid symbol. For example, we could take a random translation from ``A'', ``U'', ``C''', ``G'' to  ``H'', ``E'', ``T''', ``L'', to get the following translation:
\begin{align*}
    &\text{``A'' ( Adenine)} \rightarrow \text{``H'' (Histdine)} \\
    &\text{``U'' (Uracil)}   \rightarrow \text{``E'' (Glutamic Acid)} \\
    &\text{``C'' (Cytosine)} \rightarrow \text{``T'' (Threonine)} \\
    &\text{``G'' (Guanine)}  \rightarrow \text{``L'' (Leucine)}
\end{align*}

As we show in our experiments, a random translation between nucleotide and amino acid symbols would be sufficient for adapting the protein contact prediction model, CoT, to RNA contact prediction. However, as discussed in Section~\ref{section:different-translations}, smarter translations, which can be manually devised or learned, could result in a better performance for our adapted RNA contacted prediction model. 

\subsubsection{The Learnable Model}\label{Methods:model}
The CoT model has six consecutive attention blocks and one refinement block, each outputting a $L \times L \times C$ attentive feature map, where $C$ is a hyper-parameter corresponding to the number of features being learned by the model. In the original implementation of the protein CoT, $C$ is set to $96$, and the output of each attention block (i.e., the feature map for that block) is fed into the next block (see the ``Adapted Feature Extraction'' row in Figure \ref{fig:overview}).

For our RNA contact prediction, we further attach four layers of 2D convolution (Conv2d) modules to the intermediate feature map outputs for each of the seven attention blocks described above (see the ``Feature Fusion'' row in Figure \ref{fig:overview}). We concatenate the output of the Conv2d modules for each of the seven attention blocks into one $L \times L \times (C \times 7)$ tensor and finally pass it to a classifier module with 3 Conv2d layers for contact prediction (see the ``classification'' row in Figure \ref{fig:overview}). The output of our model has shape $L \times L \times 37$, i.e., the distance between pairs of nucleotides is divided into 37 bins. Our model is trained using standard cross-entropy loss with the bins as labels. The summed probability value of the bins for a distance less than 10\r{A} is used as the final contact prediction.

The complete architecture of our model is visualized in Figure~\ref{fig:overview}.

\subsection{Dataset} 
We use a publicly-available well-curated RNA dataset used by \citet{zerihun2021coconet}. We use the provided data split for training, validation, and testing. \texttt{RNA\_DATASET} is used for training (and validation) and \texttt{RNA\_TESTSET} is for testing. We removed 3 RNAs (RF02540, RF01998 and RF02012) whose sequences are too long for CoT. In total, we have 56 RNAs for training and validation and 23 RNAs for testing, all from different RNA families. We set the maximum number of homology sequences in MSA to be 200, based on the limits of our GPU memory. This constraint can be alleviated if large GPUs are available.

\subsection{Baseline Methods}
We compare our method to several representative MSA-based methods: (1) Unsupervised methods: mfDCA and plmDCA, using the implementations from \texttt{pydca} \cite{zerihun2020pydca}, and PSICOV \cite{jones2012psicov}, and PLMC \cite{hopf2017mutation} (2) Supervised method: CoCoNet.

Note that all these baselines are variants of DCA or are based on DCA (the current trend in studying RNA contacts). Our proposed method does not rely on DCA and approaches the problem from a different angle through the transfer learning of learned knowledge from a pre-trained protein contact prediction model.

\subsection{Training Details}
We use the Adam optimizer \cite{kingma2015adam} and cosine anneal learning rate scheduler with an initial learning rate of $1e^{-3}$. We train on an RTX-A6000 GPU using PyTorch-1.8 and CUDA-11 and search the hyper-parameters for the total training epochs among $\{100,300,500\}$ and batch sizes among $\{4,8,12,16\}$. The code is available for review at \url{https://anonymous.4open.science/r/ICML2024-submission-792}.

We randomly divide the 56 RNAs reserved for training into 47 RNAs for training and 9 RNAs for validation and use the given 23 RNAs in the test dataset for testing. The best-validated model during the training is used for testing. See Appendix~\ref{Appendix:validation-splits} for details of the validation splits.

\subsection{Evaluation Metrics}
Following the standard protocol of prior works \cite{jian2019direct, zhang2021rnacmap, sun2021rna, singh2022predicting}, we evaluate the precision on each RNA sequence of length $L$ with top-${L}$ predictions of each method ($\text{PPV}_{L}$); i.e., for an RNA with a sequence of length $L$, we use our model to make $L$ predictions (for all different $(i,j)$ pairs). Among these $L$ predictions, if $K$ pairs are true contacts, then $\text{PPV}_{L} = \frac{K}{L}$. We also report results for $\text{PPV}_{0.3/0.5L}$.

\section{Results}

Unless specified otherwise, the results presented in this manuscript employ translation nucleotides to amino acids (AUCG $\rightarrow$  HETL) as detailed in Sec.~\ref{Methods:input}.

\subsection{Main Results}\label{section:Experiments}
We first compare our method to those only using MSA as input to evaluate the contribution of the pre-trained protein Transformer to RNA contact prediction. Unsupervised algorithms like mfDCA, plmDCA, PSICOV, and PLMC are based on covariance analysis. CoCoNet uses DCA output as input and ground truth contact maps to train a supervised ConvNet classifier. We compare six CoCoNet configurations from \citet{zerihun2021coconet}.

Table~\ref{table:top-L-PPV} shows supervised models significantly outperform unsupervised baselines. Our transfer learning-based model outperforms the best CoCoNet configuration by an absolute of $5.0$, $7.4$, and $7.8$ for $\text{PPV}{L}$, $\text{PPV}{0.5L}$, and $\text{PPV}_{0.3L}$, respectively (see Appendix~\ref{Appendix:visual-comparison} for contact visualization).

\begin{table}[!t]
\begin{center}
\small
\begin{tabular}{l l l l}
           \hline
           Method & $\text{PPV}_{L}$ & $\text{PPV}_{0.5L}$ & $\text{PPV}_{0.3L}$ \\
           \hline
           \hline
           mfDCA  & 34.1 & 46.7 & 57.4 \\
           plmDCA & 30.6 & 43.2 & 57.8\\
           PSICOV & 32.1 & 43.8 & 57.8 \\
           PLMC   & 33.5 & 45.9 & 57.4 \\
           CoCoNet$^{\mathsection}$ $_{(3 \times 3)}$ & 61.6 & 67.7 & 69.1 \\
           CoCoNet$^{\mathsection}$ $_{(5 \times 5)}$ & 61.8 & 65.2 & 67.8\\
           CoCoNet$^{\mathsection}$ $_{(7 \times 7)}$ & 62.4 & 66.6 & 69.2 \\
           CoCoNet$^{\mathsection\mathparagraph}$ $_{(3 \times 3) \times 2}$ & 67.1 & 71.6 & 72.3\\
           CoCoNet$^{\mathsection\mathparagraph}$ $_{(5 \times 5) \times 2}$ & 67.5 & 71.9 & 75.0\\
           CoCoNet$^{\mathsection\mathparagraph}$ $_{(7 \times 7) \times 2}$ & 68.5 & 73.2 & 75.2\\
           \hline
           CoT-RNA$^{\dagger}$ (Ours) &\textbf{73.5} & \textbf{80.6} & \textbf{83.0} \\
           CoT-RNA$^{\ddagger}$ (average)& \textbf{72.1} $_{\pm 1.1}$ & \textbf{79.2} $_{\pm 1.7}$ & \textbf{81.9} $_{\pm 2.2}$
           \\
           \hline
\end{tabular}
\caption{Comparison of different RNA contact prediction methods based on MSA. $\mathsection$: Using the publicly released parameters by \citet{zerihun2021coconet}, which are trained using our training and validation sets. $\mathparagraph$: Using prior knowledge of Watson-Crick pairs is used. $\dagger$: Our models trained using only the training set, selected based on the best validation and evaluated on the testing set. $\ddagger$: We repeat the experiments four times using different random training and validation splits and report the mean and standard deviation of the results on the test dataset.}
\label{table:top-L-PPV}
\end{center}
\end{table}

CoCoNet is designed to be shallow, with a few parameters to learn, given the very limited number of available RNA contact prediction training data and features (from DCA). In contrast, by transfer learning of CoT, a large pre-trained protein contact prediction model, our model is much larger and deeper and can learn more diverse features through the multi-layer attentions of CoT contain. To investigate whether the improvement of our model over CoCoNet is mainly based on its capacity or the learned knowledge of the pre-trained protein language-model, we also implement a Deep-CoCoNet, which takes the same inputs as CoCoNet but replaces the CoCoNet's original shallow ConvNet with a deep model similar to ours (with the same number of layers and each layer with the same number of channels, except for the first input layer). We find that Deep-CoCoNet performs much worse than the original shallow CoCoNet, possibly due to the hardness of fitting a large model that uses features with limited expressiveness (this may also explain why \citet{zerihun2021coconet} use a shallow convolution network over deeper ones in their implementation). 

While CoCoNet takes DCA as input which is a tensor of $1 \times L \times L$ ($L$ being the RNA sequence length), our method, leveraging multi-layer CoT, has diverse attentive features of shape $(7 \times 96) \times L \times L$ (There are $7$ layers in CoT and each layer outputs $96$ channels/features). These diverse features allow us to learn deeper and larger models that can better generalize. In Section~\ref{Ablations:different-model-size}, we further investigate this, showing that our model indeed benefits from increasing the number of parameters in the transfer modules.

Additionally, we compare our methods with hybrid approaches and RNA secondary predictions in Appendix~\ref{Appendix:hybrid-approaches} and Appendix~\ref{Appendix:secondary-structure}, respectively.

\subsection{Ablation Studies}\label{section:Ablations}
Here, we examine the design choices of our model for transfer learning by exploring different configurations.

\begin{figure}[!t]
\centering
\includegraphics[width=0.45\textwidth]{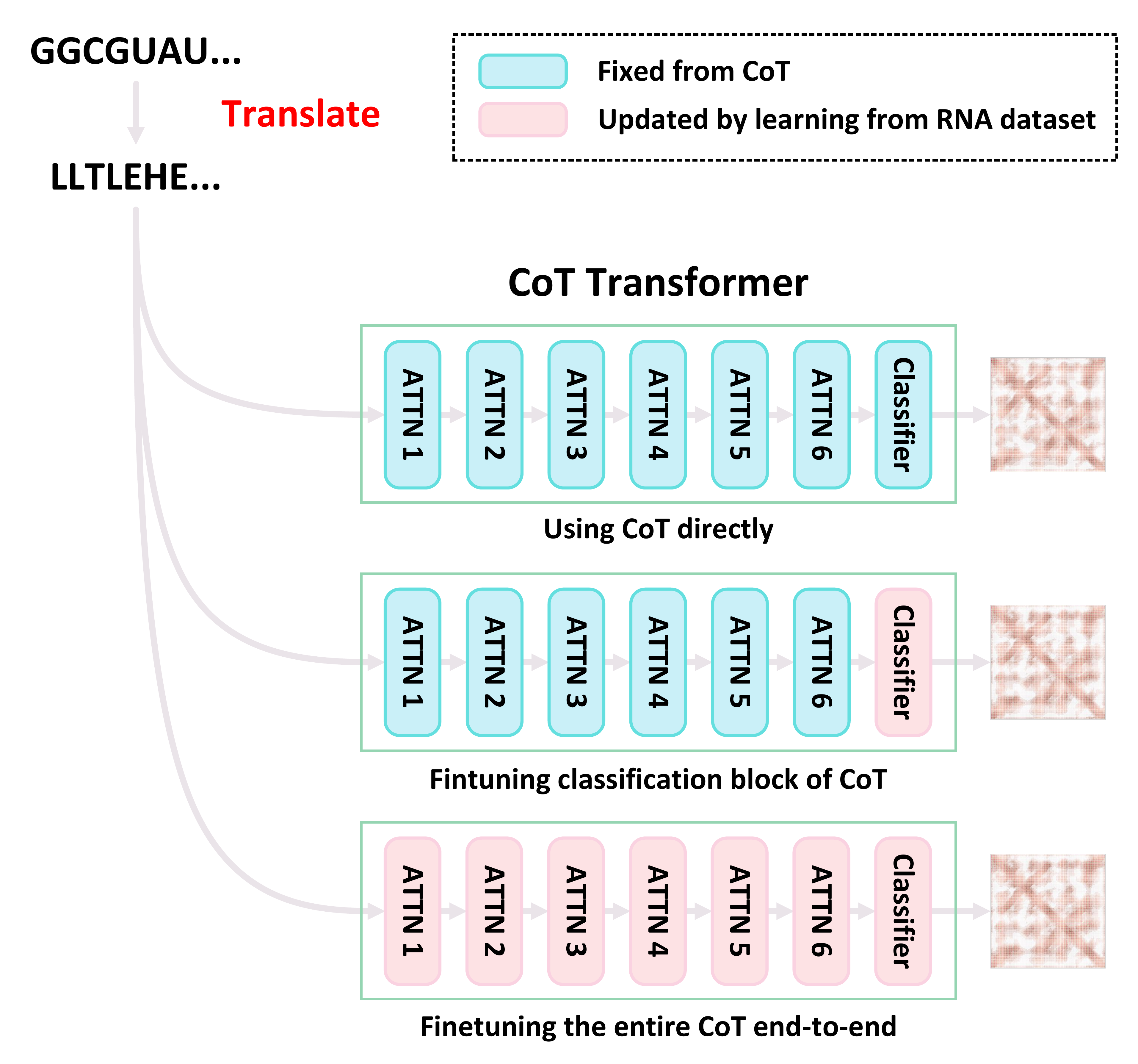}
\caption{Common baselines for transferring protein CoT to RNA contact prediction.}
\label{fig:what-to-transfer}
\end{figure}

\subsubsection{Common Transfer Learning Strategies}
We investigate three common transfer learning strategies (see Figure~\ref{fig:what-to-transfer} for an overview) and show that they are not well-suited for this task:

\textbf{Using CoT directly (\textbf{CoT directly}).} By adapting the embedding to map RNA nucleotides to protein amino acids, the pre-trained CoT can directly output a prediction of a distance map for RNA contact prediction without any model modifications.

\textbf{Fine-tuning the classification block of CoT (\textbf{CoT cls fine-tuned}).} A typical approach in transfer learning is to fine-tune the last few layers of a model. CoT has six attention blocks followed by a final ResNet block for prediction. We attach a new classification block while keeping others fixed.

\textbf{Fine-tuning the entire CoT end-to-end (\textbf{CoT end-to-end}).} Another common protocol for transfer learning is to fine-tune the entire pre-trained model end-to-end. We update all parameters in the pre-trained protein CoT by the RNA training set.

Table~\ref{table:ablation-transfer} shows the performance of these methods on our dataset. With a $\text{PPV}_{L}$ of $30.4$, the direct use of CoT (CoT directly) without any learning is shown to be inefficient. This suggests that learned protein knowledge by itself cannot be successfully transferred to RNA tasks without some fine-tuning. The results for transfer learning through fine-tuning the classification block of CoT (CoT cls fine-tuned) are considerably better, being competitive with the mfDCA baseline. These results suggest that tuning the attention features in the last layer of CoT enables the transfer of knowledge to the RNA tasks to some extent. However, as this configuration ignores the attention features from the other layers, it performs significantly worse than our method, suggesting that these features also play an important role in contact prediction and need to be tuned for RNA contact prediction. Finally, the end-to-end model, which updates all the parameters in CoT (CoT end-to-end), performs similarly to the last variant. Though this model does not ignore any part of the CoT, it requires the tuning of $\sim$20M parameters. With only 56 RNA training points, the model is likely to over-fit.

From these experiments, we can conclude that the effective transfer of protein CoT to the RNA contact prediction task requires (1) adapting some of the parameters of CoT to the new task, (2) leveraging the multi-layer attention features from CoT, and (3) making the number of learnable parameters ``small'' and proportional to the size of the training set for the new task. These findings lead to our final model design described in Section~\ref{Methods:model} that leverages multi-layer attention features from CoT and learns an appropriate number of parameters for our small training set.

\begin{table}[!t]
\begin{center}
\small
\begin{tabular}{l c c c}
           \hline
           Method & $\text{PPV}_{L}$ & $\text{PPV}_{0.5L}$ & $\text{PPV}_{0.3L}$ \\
           \hline
           mfDCA (baseline) & 34.1 & 46.7 & 57.4 \\
           CoT directly &  30.4 & 33.1 & 34.0 \\
           CoT cls fine-tuned & 38.3 & 41.6 & 41.2 \\ 
           CoT end-to-end    & 36.2 & 43.2 & 46.6 \\
           Ours & \textbf{73.5} & \textbf{80.6} & \textbf{83.0} \\
           \hline
\end{tabular}
\caption{\small{Common transfer learning strategies applied to CoT.}}
\label{table:ablation-transfer}
\end{center}
\end{table}

\subsubsection{Feature Fusion Design Choices}\label{Ablations:fusion-design}

\begin{figure}[!ht]
\centering
\includegraphics[width=0.47\textwidth]{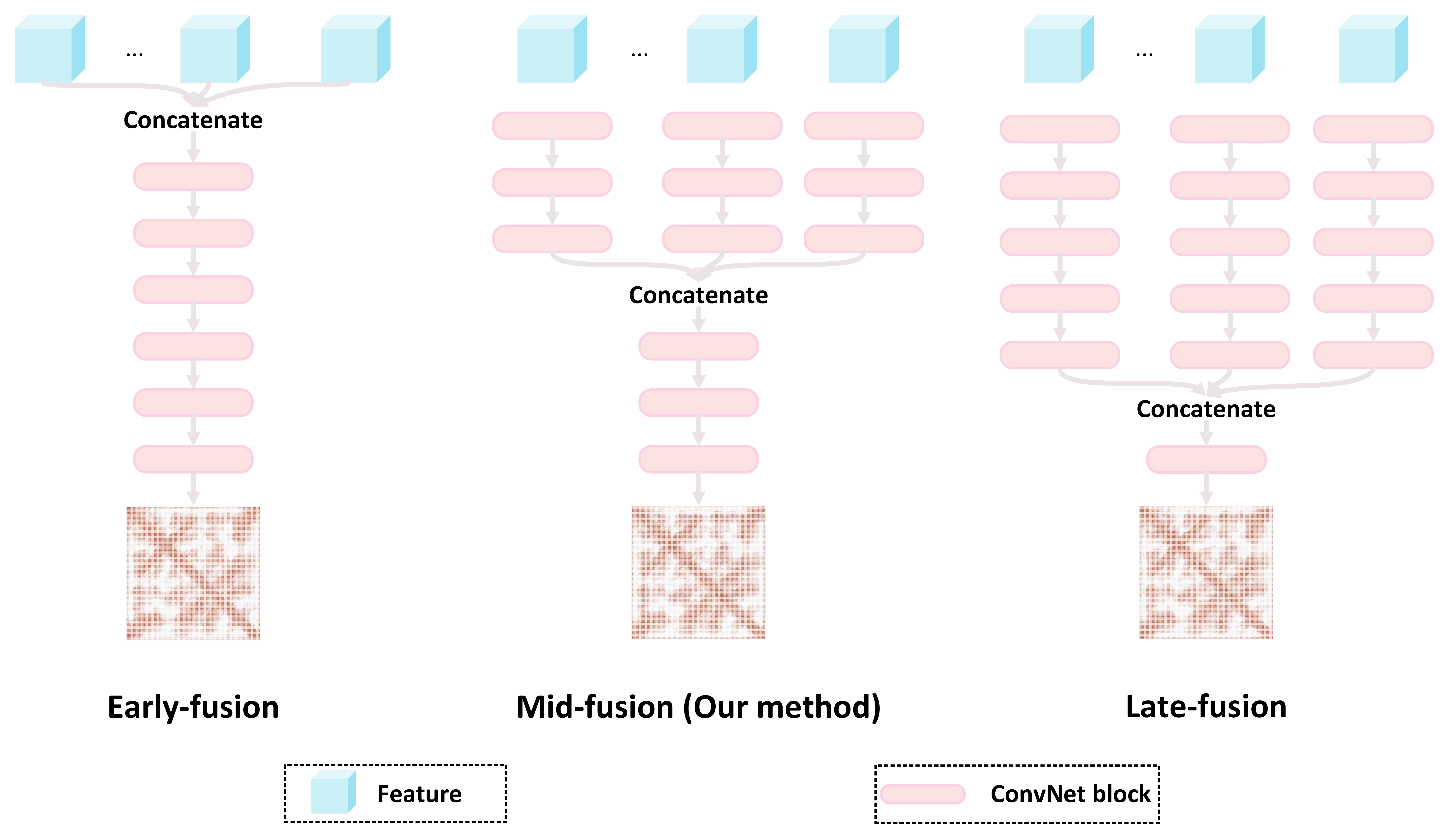}
\caption{Different feature fusion strategies. Our final model (shown in Figure~\ref{fig:overview}) uses the \emph{mid-fusion} design.}
\label{fig:fusion-styles}
\end{figure}

We examine various strategies for combining attention features from different CoT layers/blocks. Our model, as shown in Figure~\ref{fig:overview}, employs multi-branch networks (each with 4 ConvNet layers) followed by a shared 3-layer ConvNet classification block. Each branch network separately processes the attention features of CoT at each layer, before being fused and passed into the classification block (termed \emph{mid-fusion} design). Other designs include \emph{early-fusion} and \emph{late-fusion}. Early-fusion concatenates all CoT features from different layers and processes them using a single shared network. Late-fusion has separate branch networks for each CoT layer's features before being merged at the very end, followed by a single classification layer. Figure~\ref{fig:fusion-styles} provides a schematic diagram of these three fusion strategies.

\begin{table}[!t]
\begin{center}
\small
\begin{tabular}{l c c c}
           \hline
           Method & $\text{PPV}_{L}$ & $\text{PPV}_{0.5L}$ & $\text{PPV}_{0.3L}$ \\
           \hline
           early-fusion & 71.8 & 79.5 & 82.4 \\
           mid-fusion (Ours)  & \textbf{73.5} & \textbf{80.6} & \textbf{83.0} \\
           late-fusion & 72.0 & 77.7 & 79.4 \\
           \hline
\end{tabular}
\caption{\small{Comparison of different feature fusion designs. We modify the number of channels in each layer so that all three models have a similar number of parameters for fair comparison. }}
\label{table:ablation-fusion}
\end{center}
\end{table}

To make the comparison of these three designs fair, we modify the number of channels in each layer so that the three models have a similar number of parameters. As shown in Table~\ref{table:ablation-fusion}, while all design choices work well, \emph{mid-fusion} has the best performance. It is possible that features from different layers contain different types of information that may need to be processed by different ``expert'' models (i.e., ConvNet branches in our model), making an early-fusion model inefficient. In both \emph{mid-fusion} and \emph{late-fusion}, each branch network will process the attention features of each layer separately, with \emph{late-fusion} having a relatively smaller classification head. The overall better performance of \emph{mid-fusion} suggests that a good design choice is to have a balanced distribution of parameters into the branch networks and the classification head.

These experiments provide insights into the design of hybrid RNA contact prediction methods \cite{zhang2021rnacmap, sun2021rna, singh2022predicting}, which all adopt the \emph{early-fusion} design. We further discuss this in Section~\ref{section:Discussion}.

\subsubsection{Different Model Sizes}\label{Ablations:different-model-size}
As discussed in Section~\ref{section:Experiments}, the deep-CoCoNet variant of CoCoNet under-performs compared to the shallower original CoCoNet, likely due to the limited expressiveness of input DCA features which are single channel with a shape of $[1 \times L \times L]$. In contrast, here we demonstrate that our transferred CoT model allows for learning deeper networks, with its performance improving as we increase the parameters in the transfer modules.

We create larger and smaller versions of our model by increasing and decreasing the number of channels in each layer, respectively. As shown in Table~\ref{table:ablation-size}, the larger models outperform smaller ones, possibly due to the expressiveness of the CoT features from the 7 different attention blocks.

\begin{table}[!t]
\begin{center}
\small
\begin{tabular}{l c c c}
           \hline
           Method & $\text{PPV}_{L}$ & $\text{PPV}_{0.5L}$ & $\text{PPV}_{0.3L}$ \\
           \hline
           Ours (small) & 71.1 & 76.0 & 79.0 \\
           Ours & 73.5 & 80.6 & 83.0 \\
           Ours (large) & \textbf{78.3} & \textbf{82.3} & \textbf{86.1}
           \\ 
           \hline

\end{tabular}
\caption{\small{Comparison of our model with different sizes. While maintaining the network structures, we vary the number of channels in each layer so that we end up with models with different numbers of parameters.}}
\label{table:ablation-size}
\end{center}
\end{table}

\subsubsection{Protein to RNA Translation Variations}\label{section:different-translations}
We have used a random translation from RNA nucleotides to protein amino acids (e.g., ``AUCG'' to ``HETL'') in our experiments. Here, we study the effects of different translations on our model's performance.

The 20 amino acids can be categorized into four groups: (1) electrically charged, (2) polar uncharged, (3) hydrophobic, and (4) special cases. Randomly selecting one from each group generally works well (e.g., ``AUCG'' $\rightarrow$ ``RDSY'' and ``AUCG'' $\rightarrow$ ``KDNY'' in Table~\ref{table:ablation-translation}), indicating our framework's robustness to translation choices.

We also test possibly one of the worst translations, ``AUCG'' $\rightarrow$ ACDE'', as it may generate unlikely amino acid chains (e.g., a string of negatively charged residues, as ``D'' and ``E'' are negatively charged) and hence CoT will have had limited exposure to such sequences during its pre-training. Though we see a relative performance drop, the results are still comparable to CoCoNet. Appendix~\ref{Appendix:more-translations} offers a more comprehensive analysis, demonstrating our method's robustness to various translations.  Although the choice of amino acids may slightly impact performance, permutations of amino acid sets yield similar outcomes.

\begin{table}[!t]
\begin{center}
\small
\begin{tabular}{l c c c}
           \hline
           Method & $\text{PPV}_{L}$ & $\text{PPV}_{0.5L}$ & $\text{PPV}_{0.3L}$ \\
           \hline
           AUCG $\rightarrow$ ACDE & 68.6 & 76.9 & 82.5  \\
           AUCG $\rightarrow$ HETL & \underline{73.5} & \underline{80.6} & \underline{83.0} \\
           AUCG $\rightarrow$ RDSY & 76.1 & 81.4 & 83.4 \\
           AUCG $\rightarrow$ KDNY & \textbf{77.3} & \textbf{84.5} &  \textbf{88.6}
           \\ \hline
           
\end{tabular}
\caption{\small{Results of different translations/translations from nucleotides to amino acids of our transferred CoT. Bold corresponds to the best-performing translation; underline corresponds to the main translation used in the experiments.}}
\label{table:ablation-translation}
\end{center}
\end{table}

A learnable $4 \times 20$ nucleotide-to-amino acid embedding could yield better results but faces implementation challenges, such as requiring powerful GPUs and adapting the original CoT model's separate binary executable embedding layer to the PyTorch framework.

\section{Discussion and Conclusion}\label{section:Discussion}
We demonstrate the effectiveness of transferring CoT, a pre-trained protein Transformer model for contact prediction, to the RNA contact prediction task using a small curated RNA dataset. Unlike hybrid methods, our approach does not use additional features extracted by RNA analysis tools (e.g., RNAcontact). Incorporating CoT features and RNA features (extracted by tools like RNAcontact) could potentially improve our method's performance.

Note that even though our method uses a Transformer-based architecture that models attention between every element in a sequence, the maximum sequence length is typically limited due to computational constraints to a few hundred elements. Also, we only sample 200 homologous sequences from the multiple sequence alignment (MSA) of an RNA as input. This sampling process may result in the loss of co-evolutionary information, thus limiting the learning capacity of CoT. Furthermore, instead of utilizing a ``manual translation'' of CoT with a translation of AUCG $\rightarrow$ HETL, a ``soft-learnable translation'' from RNA to proteins (a 4$\times$20 matrix) could potentially yield improved results. Nevertheless, as discussed in Sec.~\ref{section:different-translations}, implementing such an approach currently faces several engineering challenges.

Our ablation experiments on model size suggest that using a larger model and modestly increasing the training set size may improve predictions. Combining these insights could yield a powerful RNA contact prediction model without requiring large-scale RNA data curation.

Our feature fusion experiments show that mid-fusion outperforms early-fusion and late-fusion. Current hybrid RNA, contact prediction tools, adopt an early-fusion design \cite{zhang2021rnacmap, sun2021rna, singh2022predicting}, concatenating DCA features and RNA features into a single tensor, followed by learning a Deep ResNet or ConvNet. Intuitively, features from different RNA tools capture distinct knowledge. Therefore, employing dedicated modules for processing features from each tool and using a mid-to-late fusion strategy could improve hybrid method performance.

Our findings shed light on a compelling representation transfer problem in computational structural biology; specifically, we investigate if structural patterns learned from large-scale protein datasets can be transferred to data-scarce RNA problems, particularly for structural contact predictions. Our results indicate that protein-to-RNA transfer learning can improve RNA model performance, suggesting that other pre-trained protein Transformers, such as MSA-Transformer \cite{rao2021msa} and ESM \cite{rao2021transformer}, could potentially be transferred to RNA for other downstream tasks.


\section{Limitations}

Note that even though our method uses a Transformer-based architecture that models attention between every element in a sequence, the maximum sequence length is typically limited due to computational constraints to a few hundred elements. Also, we only sample 200 homologous sequences from the multiple sequence alignment (MSA) of an RNA as input. This sampling process may result in the loss of co-evolutionary information, thus limiting the learning capacity of CoT. Furthermore, instead of utilizing a ``manual translation'' of CoT with a translation of AUCG $\rightarrow$ HETL, a ``soft-learnable translation'' from RNA to proteins (a 4$\times$20 matrix) could potentially yield improved results. Nevertheless, as discussed in Sec.~\ref{section:different-translations}, implementing such an approach currently faces several engineering challenges.

\section*{Impact Statements}
The primary contribution of this study lies in demonstrating the transferability of structural patterns derived from pre-trained protein Transformers to RNA tasks. Specifically, we introduce an algorithm for RNA contact prediction. Our proposed method for RNA contact prediction holds significant promise in advancing the field of RNA 3D structural predictions. This advancement, in turn, has far-reaching implications, including its application in drug design and related areas. However, it is crucial to acknowledge the potential limitations of our method. False predictions made by our algorithm could result in inaccuracies in RNA structure predictions, thereby exerting a detrimental influence on subsequent downstream tasks. 

\bibliography{example_paper}
\bibliographystyle{icml2024}

\newpage
\appendix
\onecolumn

\section{Validation Splits}\label{Appendix:validation-splits}

As explained in Section~\ref{section:experimental-setup}, we randomly divide the 56 training RNAs (i.e., \texttt{RNA\_DATASET} ) into 47 RNAs for training and 9 RNAs for validation. We use the provided 23 test RNAs (i.e., \texttt{RNA\_TESTSET}) for testing. The best validated model during training is used for testing.  

We provide the validation splits in Table~\ref{table:validation-splits}. Most experiments are carried out using Validation Set 1, with the exception that we average the results of the four different runs using all validation splits and report the means and standard deviations in Table~\ref{table:top-L-PPV}, in the main paper.

\setcounter{table}{0}
\renewcommand\thetable{A.\arabic{table}}
\begin{table}[ht]
\begin{center}
\begin{tabular}{c c c c}
           \hline
           $\text{Validation Set 1}$ & $\text{Validation Set 2}$ & $\text{Validation Set 3}$ & $\text{Validation Set 4}$ \\
           \hline
           \hline
           RF01510 &  RF01826 & RF00442\_1 & RF00921 \\
           RF01689 & RF01831\_1 & RF00458 & RF01051 \\ 
           RF01725    & RF01852 & RF00504 & RF01054 \\
           RF01734 & RF01854 & RF00606\_1 & RF01300 \\
           RF01750  & RF01982 & RF01750 & RF01415
           \\
           RF01763  & RF02001\_2 & RF01763 & RF01510
           \\
           RF01767  & RF02266 & RF01767 & RF01689
           \\
           RF01786  & RF02447 & RF01786 & RF01725
           \\
           RF01807  & RF02553 & RF01807 & RF01734
           \\
           \hline
\end{tabular}
\caption{Validation set partitions.}
\label{table:validation-splits}
\end{center}
\end{table}

\section{Visual Comparisons Between CoCoNet and Ours}\label{Appendix:visual-comparison}
\setcounter{figure}{0}
\renewcommand\thefigure{\Alph{section}.\arabic{figure}}

\begin{figure*}[!ht]
\centering
\includegraphics[width=0.99\textwidth]{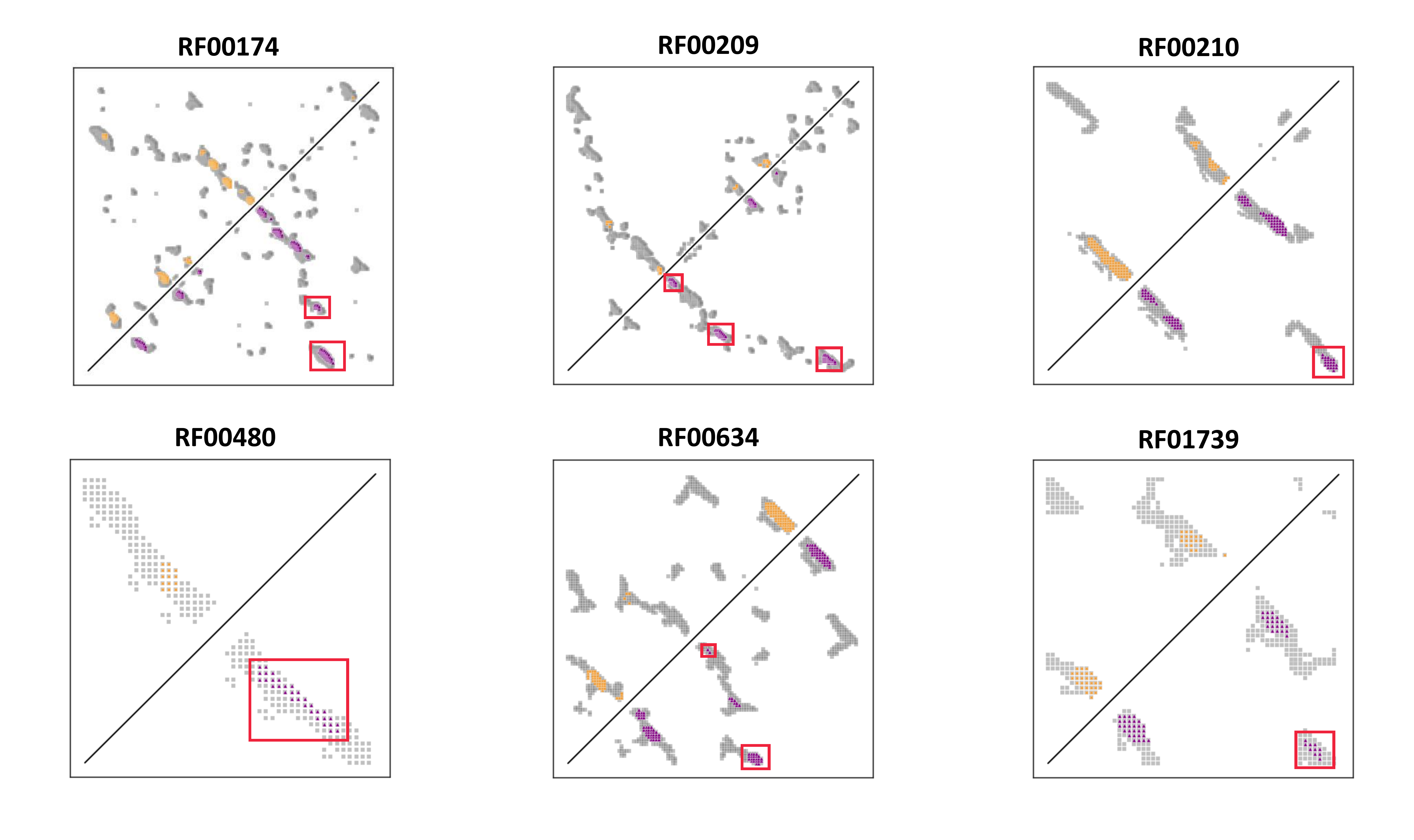}
\caption{Visual comparison of contact map predictions by CoCoNet and ours in 6 sample test RNAs. The grey dots denote ground truth contacts, the yellow dots in the upper triangle denote correct predictions by CoCoNet, and the purple dots in the lower triangle denote the correct predictions by ours. The red box highlights correct predictions by our model that are missed by CoCoNet.}
\label{fig:2d_visualization}
\end{figure*}

In this section, we present six visual examples comparing the contacts predicted by CoCoNet and those predicted by our method. Intuitively, our improved RNA contact predictions, as depicted in Figure~\ref{fig:2d_visualization}, will provide better guidance for downstream tasks.

\setcounter{figure}{0}
\renewcommand\thefigure{D.\arabic{figure}}
\begin{figure*}[!ht]
\centering
\includegraphics[width=0.99\textwidth]{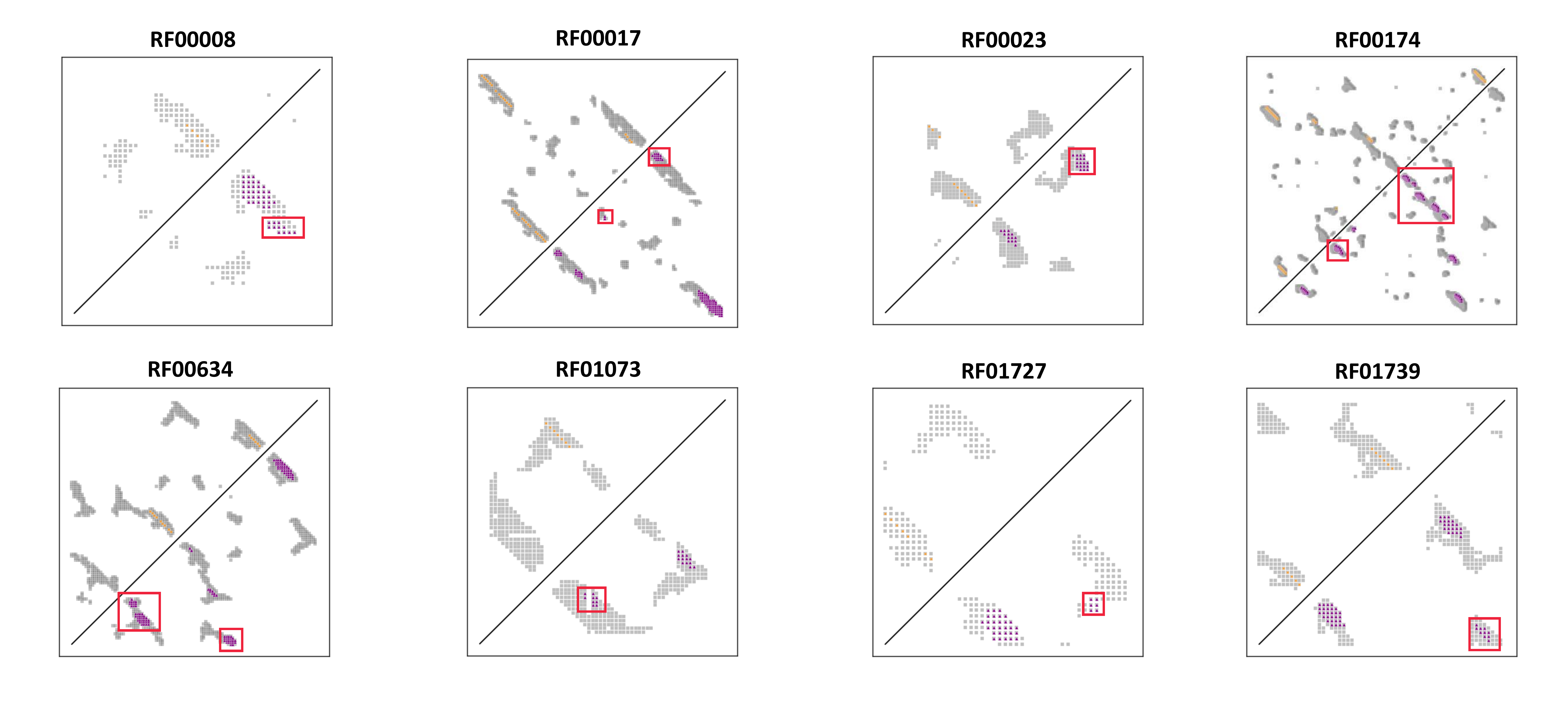}
\caption{Visual comparisons between the predictions made by LinearFold and our method. The grey dots represent the ground truth contacts, the yellow dots in the upper triangle represent the correct predictions made by LinearFold, and the purple dots in the lower triangle represent the correct predictions made by our method. The red boxes highlight the correct predictions made by our model that LinearFold missed.}
\label{fig:secondary-predictions}
\end{figure*}

\section{Comparison with Advanced Hybrid Approaches}\label{Appendix:hybrid-approaches}
Current hybrid methods for RNA contact prediction typically combine DCA with other RNA features (extracted by various RNA analysis tools) and train a supervised classifier using convolution networks. For example, RNAcontact \cite{sun2021rna} combine DCA with three other types of RNA features: (1) Covariance (computed by Infernal, which can be viewed as a variant of DCA), (2) Secondary Structures, either given or predicted by PETFold, (3) Solvent Accessible Area provided by RNAsol. 

Running hybrid methods such as RNAcontact locally can be challenging for practitioners as they may require downloading a database and building and compiling separate 3rd-party RNA tools. This becomes challenging for practitioners. Thus, many of these methods provide a web server for use by practitioners. Our model, on the other hand, is end-to-end, only taking MSA as input and outputting the prediction in a single forward pass of the model. Thus, our model can be easily run locally using a simple script, making our work accessible to a wider range of researchers with different technical backgrounds.

\setcounter{table}{0}
\renewcommand\thetable{C.\arabic{table}}
\begin{table}[!t]
\begin{center}
\begin{tabular}{l c c c }
           \hline
           Method & $\text{PPV}_{L}$ & $\text{PPV}_{0.5L}$ & $\text{PPV}_{0.3L}$ \\
           \hline
           \hline
           RNAcontact$^{\dagger}$ & 69.0 & 81.7 & 86.7\\
           Ours & 73.5 & 80.6 & 83.0 \\
           Ours (better translation) $^{\ddagger}$  & \textbf{77.3} & \textbf{84.5} &  \textbf{88.6}
           \\ \hline
\end{tabular}

\caption{Comparison of our model with RNAcontact, a hybrid approach that uses features extracted by different RNA computational tools in addition to DCA. $^{\dagger}$ indicates results obtained using a $\sim$ 5$\times$ larger training set. $^{\ddagger}$: Stronger results by using a different embedding layer for translation nucleotides to amino acids; see additional details in Section~\ref{section:different-translations}.}
\label{table:hybrid-approaches}
\end{center}
\end{table}

Nevertheless, we compare our model to RNAcontact using predictions from its web server. Note that the RNAcontact server will search MSA databases and is likely to use many more MSA sequences compared to our setting, where we use 200 MSA sequences. Moreover, RNAcontact is trained using a much larger dataset with about 300 unique target RNAs, whereas our training set only contains 56 RNAs. However, even with these advantages in favor of RNAcontact, our model still outperforms this SoTA hybrid method (see Table~\ref{table:hybrid-approaches}).

\section{Comparison with RNA Secondary Prediction Methods}\label{Appendix:secondary-structure}
In this study, we investigate RNA contact prediction, a conceptually related but distinct task from secondary structure prediction \cite{sato2021rna, radford2021learning, szikszai2022deep}. While secondary structure prediction focuses on identifying nucleotide base pairs connected through hydrogen bonding, contact prediction aims to determine all interactions within a specified distance threshold in the folded RNA structure. As a result, contact prediction can reveal more intricate details about the RNA tertiary structure.

Specifically, using multiple sequence alignment (MSA) for structural prediction, whether for exclusive pairings in secondary structure or non-exclusive pairings in contact structure, is primarily based on the assumption that structures are more conserved than sequences. This reflects presumably even more fundamental conservation in function. However, the function of a biomolecule is often determined by a cohort of non-exclusive pairings that form the pocket or cavity necessary for its function. Therefore, the predictive power extracted or learned from MSA can extend beyond the simpler exclusive pairings in secondary structure prediction (i.e., the standard base pairs with hydrogen bonds).

Here, we compare our contact prediction approach with two recent secondary structure prediction methods, demonstrating that contact prediction more effectively captures structural characteristics. We observe that secondary prediction methods generate an average of $0.3 L$ positive predictions. Thus, a more appropriate comparison utilizes $\text{PPV}{0.3L}$ (using $\text{PPV}{L}$ would place secondary prediction methods at a disadvantage).

\setcounter{table}{0}
\renewcommand\thetable{D.\arabic{table}}

\begin{table}[!t]
\begin{center}
\begin{tabular}{l c c c }

           \hline
           Method & $\text{PPV}_{L}$ & $\text{PPV}_{0.5L}$ & $\text{PPV}_{0.3L}^{\dagger}$ \\
           \hline
           \hline
           LinearFold-C & 24.5 & 49.0 & 78.1 \\
           LinearFold-V & 25.9 & 51.8 & 79.6 \\
           MXfold2      & 25.7 & 51.7 & 83.6 \\
           Ours (KDNY translation) & \textbf{77.3} & \textbf{84.5} &  \textbf{88.6}
           \\ \hline
\end{tabular}

\caption{Quantitative comparison of predictions made by LinearFold, MXfold2, and our method. The $\dagger$ column shows the PPV for LinearFold and MXfold2, which is calculated as $\frac{1}{|\mathcal{D}|}\sum_{i \in \mathcal{D}} \frac{TP_{i}}{P_{i}}$, where $TP_{i}$ represents true positives and $P_{i}$ represents the number of predictions. It is worth noting that for LinearFold $P_{i} \approx 0.30L$ and for MXfold2 $P_{i} \approx 0.32L$, making it an appropriate comparison to ours when evaluating by $\text{PPV}_{0.3L}$.}
\label{table:secondary-methods}
\end{center}
\end{table}

\setcounter{table}{0}
\renewcommand\thetable{E.\arabic{table}}
\begin{table}[!t]
\begin{center}
\begin{tabular}{l c c c}
           \hline
           Method & $\text{PPV}_{L}$ & $\text{PPV}_{0.5L}$ & $\text{PPV}_{0.3L}$ \\
           \hline
           \hline
    \multicolumn{4}{l}{\textbf{A.} Different RNA-to-protein translations}\\
           \hline
           AUCG $\rightarrow$ AVIL & 66.9 & 76.3 &	80.8  \\
           AUCG $\rightarrow$ ILMF & 67.0 & 74.0 &	78.0  \\          
           AUCG $\rightarrow$ RDSC & 72.3 & 81.7 &	82.7  \\
           AUCG $\rightarrow$ KDSU & 73.9 & 80.4 & 81.4  \\
           AUCG $\rightarrow$ HETG & 75.1 & 81.2 &	84.4  \\
           AUCG $\rightarrow$ HDTU & 76.2 & 83.2 &	86.8  \\           
           AUCG $\rightarrow$ KEQG & 76.9 & 80.9 &	81.4  \\
           AUCG $\rightarrow$ KDTI & 76.9 & 83.6 & 85.8  \\
           AUCG $\rightarrow$ KDQG & 77.3 & 84.8 &	87.7  \\
           AUCG $\rightarrow$ KDNU & 78.2 & 86.2 &	88.0  \\
           AUCG $\rightarrow$ RSKY & 78.6 & 84.3 &	87.7  \\
           AUCG $\rightarrow$ SDYK & 79.8 & 85.5 &	87.7    
           \\ 
           \hline
           \hline
    \multicolumn{4}{l}{\textbf{B.} Different permutations of KDNY}\\
           \hline
           AUCG $\rightarrow$ KYND & 74.0 &	79.8 &	83.6  \\
           AUCG $\rightarrow$ DNYK & 77.0 &	82.7 &	85.7  \\
           AUCG $\rightarrow$ NYKD & 78.6 &	83.7 &	84.5  \\
           AUCG $\rightarrow$ YNDK & 80.6 &	85.2 &	86.2
           \\ 
           \hline
           \hline
    \multicolumn{4}{l}{\textbf{C.} Different permutations of RDSY}\\
           \hline
           AUCG $\rightarrow$ RYSD & 77.2 & 82.3 &	85.9 \\
           AUCG $\rightarrow$ YRDS & 78.9 & 86.4 &	90.7 \\
           AUCG $\rightarrow$ DRSY & 80.3 & 85.4 &	87.1 \\
           AUCG $\rightarrow$ SDYR & 81.4 & 87.2 &	88.6
           \\ 
           \hline
           \hline
    \multicolumn{4}{l}{\textbf{D.} Different permutations of ACED}\\
           \hline
           AUCG $\rightarrow$ EDAC & 64.8 & 70.3 &	71.7  \\
           AUCG $\rightarrow$ CAED & 65.3 & 72.8 &	76.0  \\
           AUCG $\rightarrow$ DECA & 66.2 &76.7 &	82.7  \\
           AUCG $\rightarrow$ ACDE & 68.6 & 76.9 & 82.5 
           \\ \hline
           
\end{tabular}
\caption{Results of different Nucleotide-to-Amino Acid Translations/Mappings for our model. \textbf{A.} Despite variations in their performances, the additional translations we explored were competitive compared to the baseline CoCoNet. \textbf{B.} We also tested various translations using permutations of KDNY, which showed good results, and \textbf{C.} RDSY, which also showed promising outcomes. \textbf{D.} However, when we used permutations of ACDE, the results were relatively less competitive.}
\label{table:appendix-translations}
\end{center}
\end{table}

The results shown in Table~\ref{table:secondary-methods} ($\text{PPV}_{0.3L}$) indicate that our method outperforms LinearFold and MXfold2. Qualitative analysis and visual comparisons can be found in Figure~\ref{fig:secondary-predictions}. By design, secondary structure prediction methods generate a limited number of (secondary) predictions, while our method captures more contacts, many of which come from tertiary contacts.

We note that MXfold2 is a machine learning method. Thus, the appropriate comparison should be performed using the same training set of ours. Nevertheless, we use the MXfold2 server based on the pre-trained model trained using more than 3000 sequences \cite{sato2021rna}, and our method shows competitive results.

\section{Additional Ablation on Different Nucleotide to Amino Acid Translations}\label{Appendix:more-translations}
This section presents a more comprehensive set of experiments on various nucleotide to amino acid translations, supplementing the results in Sec.\ref{section:different-translations}. There are approximately $20^4$ distinct nucleotide-to-amino acid translations. Although exhaustively examining all combinations is not feasible, we evaluate our method using an alternative set of randomly chosen translations. As demonstrated in Table\ref{table:appendix-translations} A, our method exhibits considerable robustness with respect to the selection of different translations.
We also observe that while selecting different amino acids (e.g., DNYK or ACDE) may lead to either strong or weak performance, the permutations of a set of amino acids generally yield similar results. For example, AUCG $\rightarrow$ DNYK, KYND, NYKD, YNDK all perform well. In contrast, AUCG $\rightarrow$ ACDE, CAED, DECA, EDAC all relatively underperform. Based on these experiments, we hypothesize that the transferable contact patterns from the protein transformer CoT to RNA tasks may not rely on an exact translation. It is possible that the richer attention information learned in CoT is preserved for specific sets of amino acids.

\end{document}